\documentclass{aa}  

\usepackage{graphicx}

\usepackage{caption}
\usepackage{subcaption}
\usepackage{siunitx}

\usepackage{txfonts}
\usepackage{color}

\begin{document}

\title{Significance of bar quenching in the global quenching of star formation}

\author{K. George\inst{1}\fnmsep\thanks{koshyastro@gmail.com}, S. Subramanian\inst{2}, K. T. Paul\inst{1}}

\institute{Department of Physics, Christ University, Bangalore, India \and Indian Institute of Astrophysics, Koramangala II Block, Bangalore, India}

  \abstract{The suppression of star formation in the inner kiloparsec regions of barred disk galaxies due to the action of bars is known as bar quenching. We investigate here the significance of bar quenching in the global quenching of star formation in the barred galaxies and their transformation to passive galaxies in the local Universe. We do this by measuring the offset of quenched barred galaxies from star-forming main sequence galaxies in the star formation rate-stellar mass plane and comparing it with the length of the bar, which is considered as a proxy of bar quenching. We constructed the star formation rate-stellar mass plane of 2885 local Universe face-on strong barred disk galaxies ($z<0.06$) identified by Galaxy Zoo. The barred disk galaxies studied here fall on the star formation main sequence relation with a significant scatter for galaxies above stellar mass 10$^{10.2}$ M$\odot$. We found that 34.97 $\%$ galaxies are within the intrinsic scatter (0.3 dex) of the main sequence relation, with a starburst population of 10.78 $\%$ (above the 0.3 dex) and a quenched population of 54.25 $\%$ (below the -0.3 dex) of the total barred disk galaxies in our sample. Significant neutral hydrogen (M$_{HI}$ >10$^{9}$ M$\odot$ with log M$_{HI}$/M$\star$ $\sim$ -1.0 to -0.5) is detected in the quenched barred galaxies with a similar gas content to that of the star-forming barred galaxies. We found that the offset of the quenched barred galaxies from the main sequence relation is not dependent on the length of the stellar bar. This implies that the bar quenching may not contribute significantly to the global quenching of star formation in barred galaxies. However,  this observed result could also be due to other factors such as the dissolution of bars over time after star formation quenching, the effect of other quenching processes acting simultaneously, and/or the effects of environment.}


\titlerunning{Significance of bar quenching in the local Universe}
\authorrunning{K. George\inst{1}}

\maketitle
%
%________________________________________________________________

\section{Introduction}

The local Universe galaxies can be separated into star forming and non-star forming as evident from the bimodal distribution of broadband optical colors and star formation rates \citep{Strateva_2001,Baldry_2004,Brinchmann_2004,Salim_2007}.
The very existence of such a bimodality along with the increase in the number density of non-star-forming (passive) galaxies imply that for a good fraction of galaxies star formation is getting quenched \citep{Bell_2004,Faber_2007}. There are several processes like feedback from an active galactic nuclei (AGN feedback), bar quenching, morphological quenching, stellar feedback, mergers, ram-pressure stripping, strangulation, and harassment that can quench ongoing star formation in galaxies \citep{Peng_2015,Man_2018}.\\

Recent observations and simulations highlight the role of stellar bars in star formation quenching \citep{Haywood_2016,Hakobyan_2016,Spinoso_2017,Khoperskov_2018,James_2018,George_2019,Newnham_2019}. The disk galaxies with a stellar bar (barred galaxies) are dynamically stable and rotationally supported, where gas is transferred from the galaxy outskirts to the inner central regions through the bar-induced torques \citep{Combes_1981,Combes_1990,Debattista_2004,Kormendy_2004,Athanassoula_2013}. This leads to an enhancement in star formation at the nuclear regions of barred galaxies and along the bar itself \citep{Athanassoula_1992, Ho_1997,Sheth_2005,Coelho_2011,Ellison_2011,Oh_2012}. However, the action of stellar bars can also facilitate suppression of star formation in the annular region between the nuclear region and the ends of the bar (a few kiloparsec, equivalent to the radial length of bar) in the central regions of barred galaxies, leading to an overall reduction in the integrated star formation rate of the galaxy \citep{Tubbs_1982,Combes_1985,Reynaud_1998,Verley_2007,Masters_2010,Masters_2012,Cheung_2013,Gavazzi_2015,Haywood_2016,James_2016,Spinoso_2017,Khoperskov_2018,James_2018}. 

Two physical mechanisms are suggested in the literature for the suppression of star formation due to the action of bars (bar quenching).  The first mechanism is through the bar induced shocks and shear that can stabilize the gas against collapse by increasing turbulence and hence inhibit star formation. This is possible since the bar collects most of  the gas inside the co-rotation radius during the formation phase. (\citealt{Tubbs_1982}; \citealt{Reynaud_1998}; \citealt{Verley_2007}; \citealt{Haywood_2016}; \citealt{Khoperskov_2018}). The second mechanism is that the bar-induced torque drives gas inflows to the center of galaxy. This enhances the nuclear star formation and makes the region close to the bar devoid of fuel for further star formation (\citealt{Combes_1985}; \citealt{Spinoso_2017}). In fact such kiloparsec-scale suppressed star formation (star formation desert) has been identified in the central regions of a sample of barred galaxies \citep{James_2009} with stellar population ages in the range of $\sim$ 250 Myr - 4 Gyr  (\citealt{James_2015,James_2016,James_2018}). However, star formation deserts with stellar population ages older than 4 Gyr were not found, indicating that these regions do not last indefinitely, implying a change in the star formation property of the galaxy. 

One of the  possibilities for the absence of star formation deserts older than 4 Gyr is that the star formation suppression in the inner kiloparsec scale propagates to the outer disk of the galaxy thereby creating a passive galaxy in line with the scenario of inside-out quenching \citep{Lin_2019}. A recent integral field spectroscopy (IFU) study of nearby disk galaxies by \cite{Guo_2019} supports an inside-out quenching scenario, and found an increase in the fraction of barred galaxies in their sample with a decrease in both the global and inner-several-kiloparsec-scale star formation. This suggests the importance of dynamical processes in the global quenching of star formation in disk galaxies. The additional quenching mechanisms responsible for the propagation of star formation suppression to outside regions are not well understood. \cite{Gavazzi_2015} consider cosmological starvation as the most probable additional quenching mechanism that transforms a barred galaxy, aided by bar quenching, into passive galaxy. However, there could be other quenching mechanisms including environmental effects \citep{Skibba_2012}.  Furthermore, the gas surface density in the outer regions of the disk galaxies is found to be low and the star formation is too inefficient to contribute significantly to the galaxy total star formation \citep{Wong_2002,Fumagalli_2008,Bigiel_2010,Bigiel_2012}. 
Thus we expect bar quenching could be the primary quenching mechanism responsible for transforming barred galaxies into passive galaxies. It is to be noted that almost 2/3 disk galaxies in the local Universe host a stellar bar \citep{Mulchaey_1997,Knapen_2000,Eskridge_2000,Kormendy_2004,Menendez_2007,Sheth_2008, Nair_2010} and hence bar quenching can lead to the buildup of passive galaxies in the Universe. 
Here we attempt to understand the significance of bar quenching in the global quenching of star formation of barred galaxies.
The length of the stellar bar, a proxy for the strength of the bar \citep{Elmegreen_2007,guo_2019}, is expected to enhance the impact of bar quenching in the galaxy \citep{Spinoso_2017,James_2018}. This is possible since star formation suppression happens over the region covered by the bar length. It is therefore prudent to study the role of bar length on global quenching using a statistically large sample of galaxies in the local Universe. This will help us to understand whether bar quenching is the primary quenching mechanism in the buildup of  passive barred galaxies. In this context we study the star formation properties and their dependence on scaled bar length (bar length scaled to the size of galaxy) for a sample of face-on strong barred galaxies in the local Universe.\\

The stellar mass (M$\star$) and star formation rate (SFR) of galaxies in the local Universe form the SFR-M$\star$ plane populated by two distinct regions for star-forming and passive  galaxies \citep{Brinchmann_2004,Salim_2007}. The star-forming galaxies are found to follow a main sequence relation with an intrinsic scatter of 0.3 dex \citep{Noeske_2007,Elbaz_2007,Daddi_2007}(See also \citealt{Matthee_2019}). The relation is observed to exist up to z$\sim$4 and explains an order in the growth of galaxies \citep{Renzini_2015,Schreiber_2015}, likely due to an interplay between the gas supply, star formation, and different feedback mechanisms that regulate star formation \citep{Dave_2011,Haas_2013,Lilly_2013,Tacchella_2016,Rodrguez_2016}. The position of galaxies with respect to the main sequence can be used to identify their state of star formation  \citep{Renzini_2015, Popesso_2019}. The galaxies above and below the intrinsic scatter of the main sequence relation can be considered as starburst and quenched populations of galaxies. The SFR-M$\star$ plane of galaxies can therefore be used to understand the quenched population of barred disk galaxies in the local Universe. Here we explore the position of  barred disk galaxies on the SFR-M$\star$ plane for the local Universe with the aim of understanding the role of bar quenching in the transformation of star-forming barred galaxies into passive ones.
Then we study the dependence of bar length (a proxy for the strength of bar quenching) on the position of quenched barred galaxies in the M$\star$-specific star formation plane (sSFR:SFR divided by M$\star$). We also use the data on the neutral gas (HI) in barred galaxies to understand whether the star-forming and quenched population of barred galaxies show any difference in HI content.
We adopt a flat Universe cosmology with $H_{\rm{o}} = 71\,\mathrm{km\,s^{-1}\,Mpc^{-1}}$, $\Omega_{\rm{M}} = 0.27$, $\Omega_{\Lambda} = 0.73$ \citep {Komatsu_2011}.\\

\begin{figure*}
\centering
\subcaptionbox{}{\includegraphics[width=0.34\textwidth]{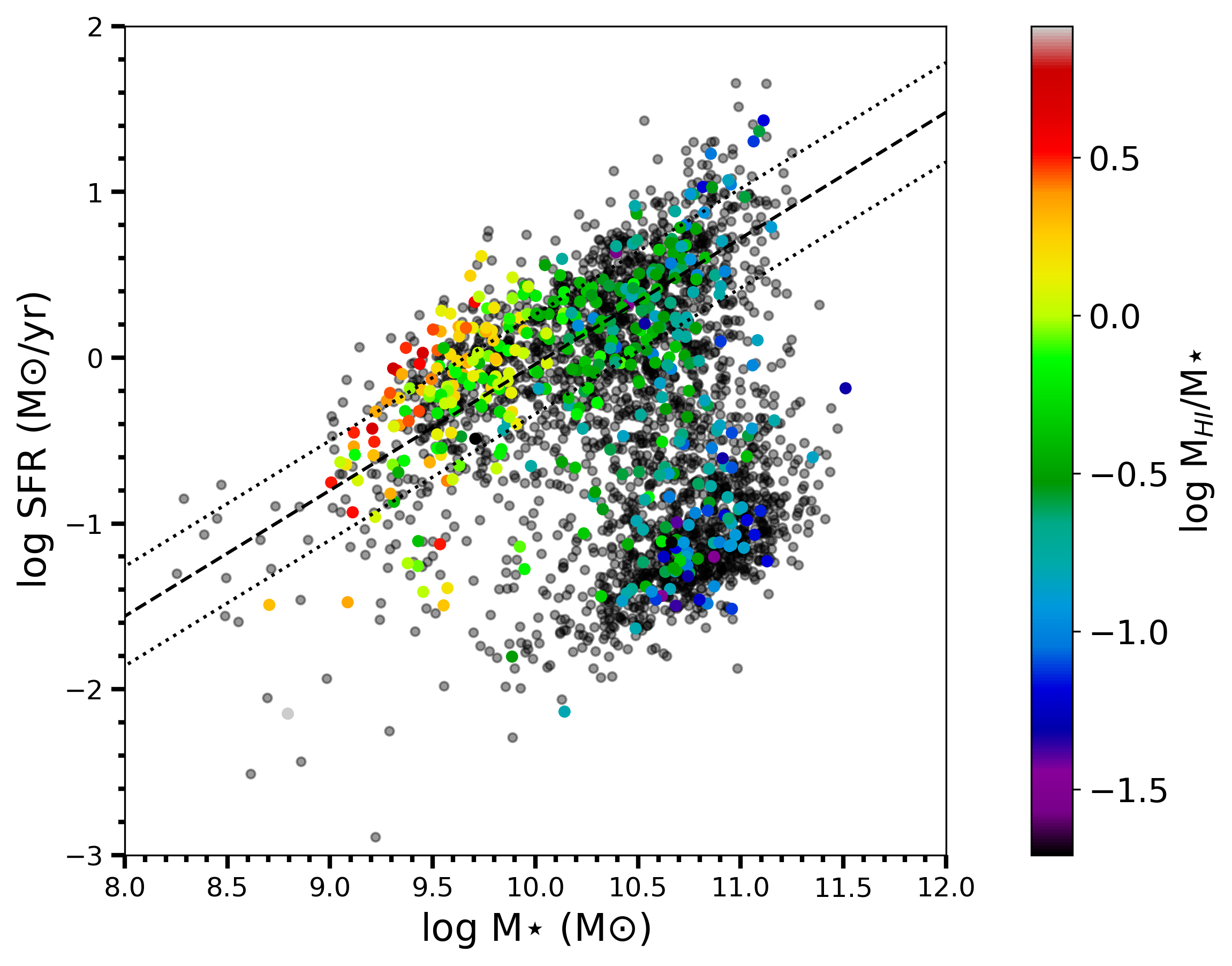}}%
%\hfill % <-- Seperation
\subcaptionbox{}{\includegraphics[width=0.32\textwidth]{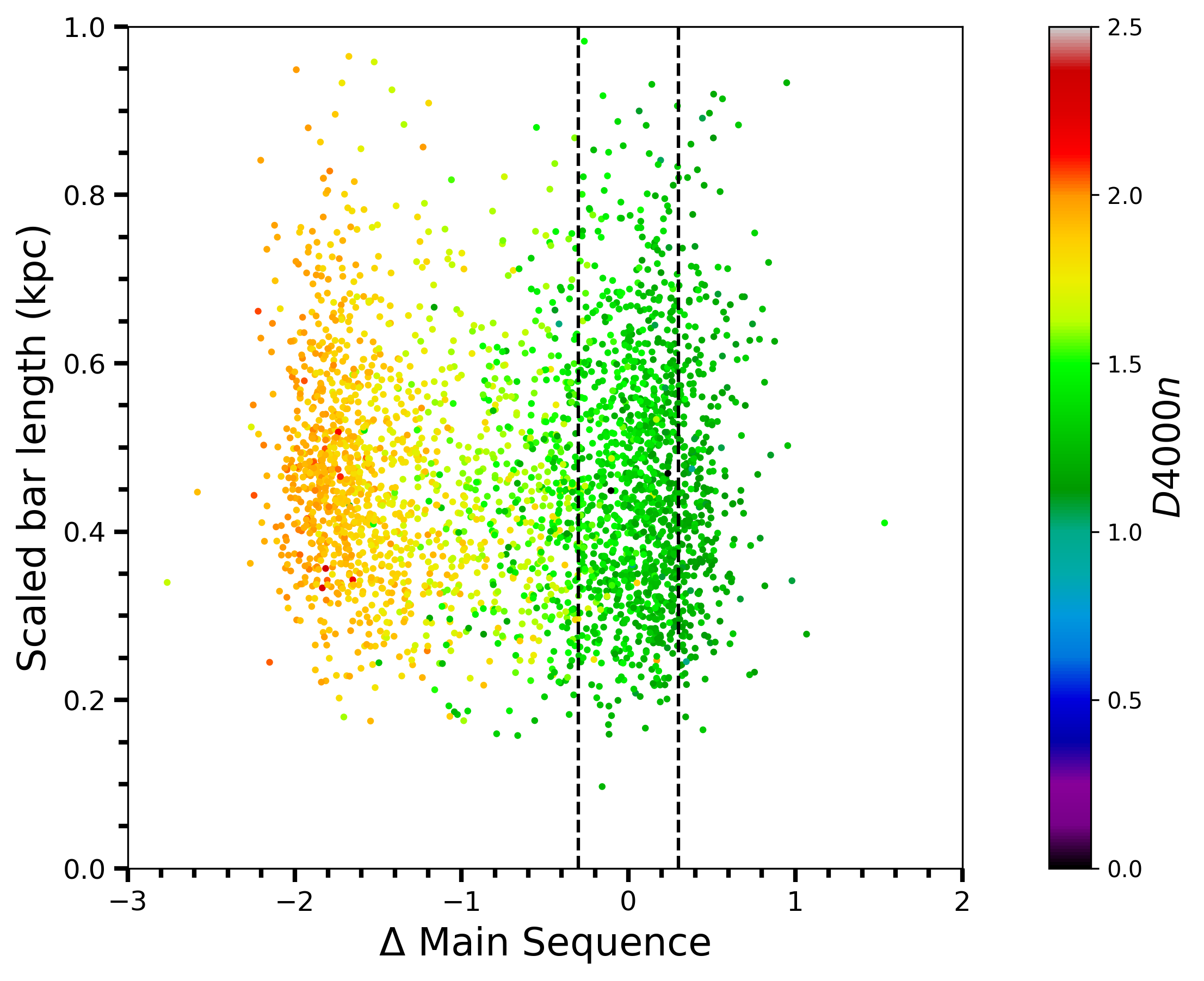}}%
%\hfill % <-- Seperation
\subcaptionbox{}{\includegraphics[width=0.34\textwidth]{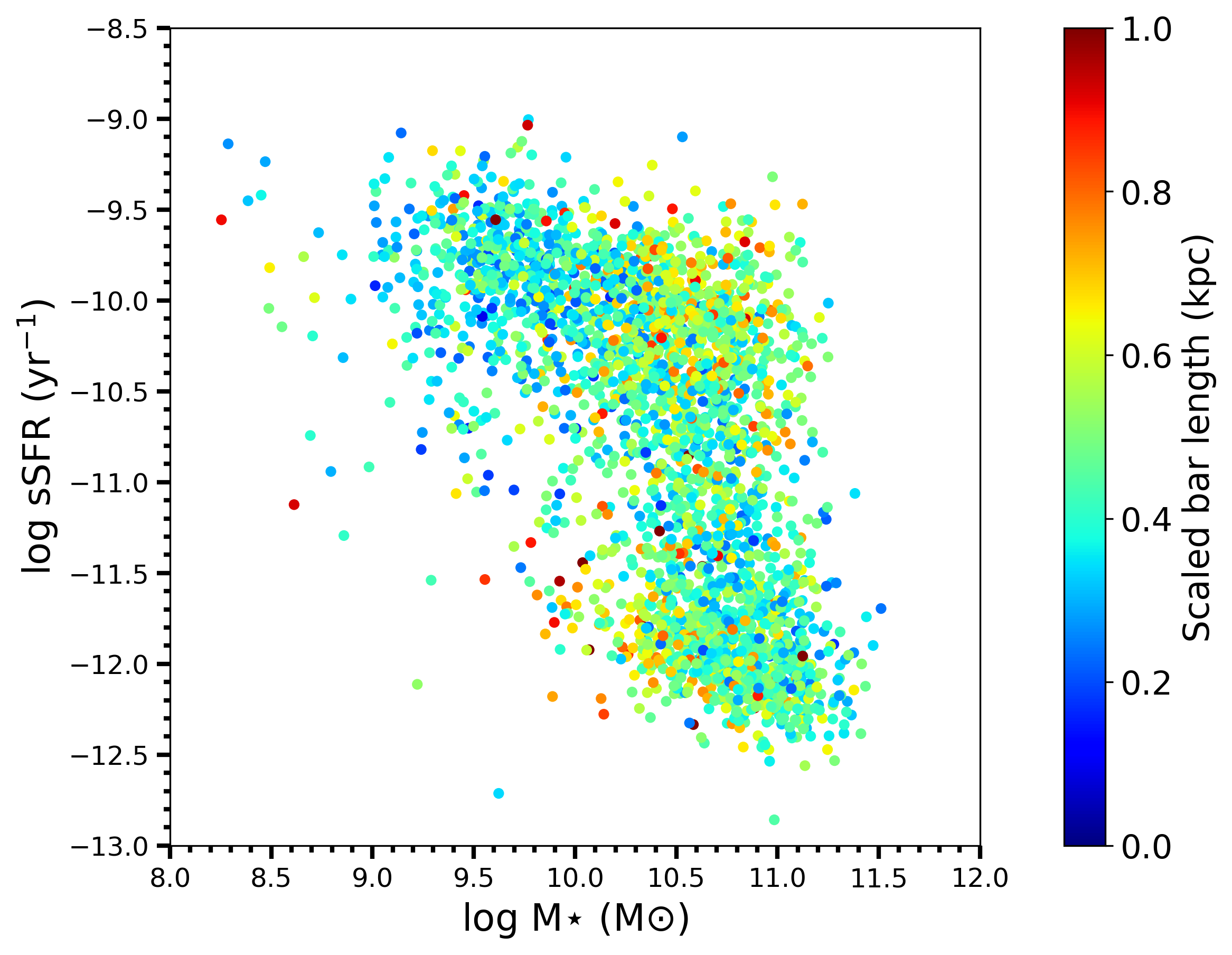}}%
\caption{SFR and M$\star$ relation (a) The SFR-M$\star$ relation for barred galaxies. The main sequence relation taken from \citet{Renzini_2015} is shown with a black dashed line and the intrinsic scatter is shown in dotted lines. The log M$_{HI}$/M$\star$ of galaxies with HI detection are shown in color scale. (b) The offset of galaxies from the SFR-M$\star$ relation is plotted against the scaled length of the bar. The $D4000n$ index strength of galaxies is shown with a color scale. (c) The sSFR-M$\star$ relation for barred galaxies. The scaled bar length is shown in color scale.}\label{figure:fig0} 
\end{figure*}

\section{Data and analysis}

The stellar bar in nearby disk galaxies can be efficiently identified from a morphological galaxy classification like the Galaxy Zoo2 (GZ2) \citep{Willett_2013}, which culminated in a catalog with details on stellar bars compiled for a list of 3150 barred galaxies, along with the spectroscopic redshift of the galaxy \citep{Hoyle_2011}. The stellar bar parameters are derived for face-on galaxies below redshift ($z$) $\sim$ 0.06. This selection criteria has the advantage of ruling out bars enshrouded in dust and the disadvantage of missing barred galaxies in highly inclined galaxies. The catalog of \citealt{Hoyle_2011} provides the scaled bar length, the absolute bar length divided by the size of the galaxy, and we use this as a proxy for the strength of bar quenching in our subsequent analysis. 
The physical parameters SFR, M$\star,$ and D4000$n$ (a proxy for recent stellar population ages)  of the sample galaxies are taken from the MPA-JHU catalog based on SDSS DR7 \footnote{https://wwwmpa.mpa-garching.mpg.de/SDSS/DR7/} \citep{Kauffmann_2003a, Brinchmann_2004} where the total SFR (corrected for extinction and aperture) computed from H$\alpha$ flux and for the case of AGN and composite galaxies is derived from the 4000 $\AA$ break strength ($D4000$) using the method described in \citealt{Kauffmann_2003a}. The total stellar mass of a galaxy is derived from fitting the broadband $ugriz$ SDSS photometry using the stellar population models of \citealt{Bruzual_2003}, assuming a \citet{Kroupa_2001} IMF. We used these values to construct the SFR-M$\star$ plane for barred galaxies in the local Universe ( z$<$0.06). Out of 3150 barred sample galaxies, we could retrieve reliable non-zero values of parameters for only 3068 galaxies. There can be contamination from higher excitation emission lines due to AGN at the center of barred galaxies that could alter our SFR and M$\star$ estimates. We used the line diagnostic classification based on the emission line kinematics results for SDSS using GANDALF (emissionLinesPort) and could retrieve the BPT classification for 3046 galaxies \citep{Baldwin_1981,Sarzi_2006}. We reject 161 galaxies classified as "Seyfert" and the following analysis is based on the remaining 2885 barred galaxies. We constructed the SFR-M$\star$ (in log scale) plane for the 2885 barred galaxies in our sample and we show this in Fig. \ref{figure:fig0}a. We used the main sequence relation for the local Universe galaxies described in \citet{Renzini_2015}, which is shown with a  black line (1$\sigma$ deviation shown with dotted line) in Fig. \ref{figure:fig0}a. We crossmatched the barred galaxies in our sample with the catalog created from the Arecibo Legacy Fast ALFA (ALFALFA) survey of galaxies in the local Universe \citep{Haynes_2018}. We used a search radius of 1$\arcsec$ between the optical counterpart coordinates of the ALFALFA catalog and the barred galaxy sample, and we retrieved HI masses for 422 galaxies. The log M$_{HI}$/M$\star$ value for these galaxies are color coded in Fig. \ref{figure:fig0}a. Figure \ref{figure:fig0}a demonstrates that a good fraction of barred galaxies follow the main sequence relation in the SFR-M$\star$ plane. The intrinsic scatter of the main sequence relation is 0.3 dex and the galaxies 1$\sigma$ above the relation can be considered as actively star forming (star burst) while the galaxies 1$\sigma$ below the relation can be considered as quenched galaxies. We computed the relative number of galaxies in three different domains (main sequence, starburst, and quenched) of the SFR-M$\star$ plane and give them in Table 1. We note that the scatter of barred galaxies off the main sequence relation is significantly above 10$^{10.2}$ M$\odot,$ which interestingly is the minimal mass (also the characteristic mass) for barred galaxies reported in \cite{Nair_2010}.\\

\begin{figure}
\centering
\includegraphics[width=6.0cm,height=6.0cm,keepaspectratio]{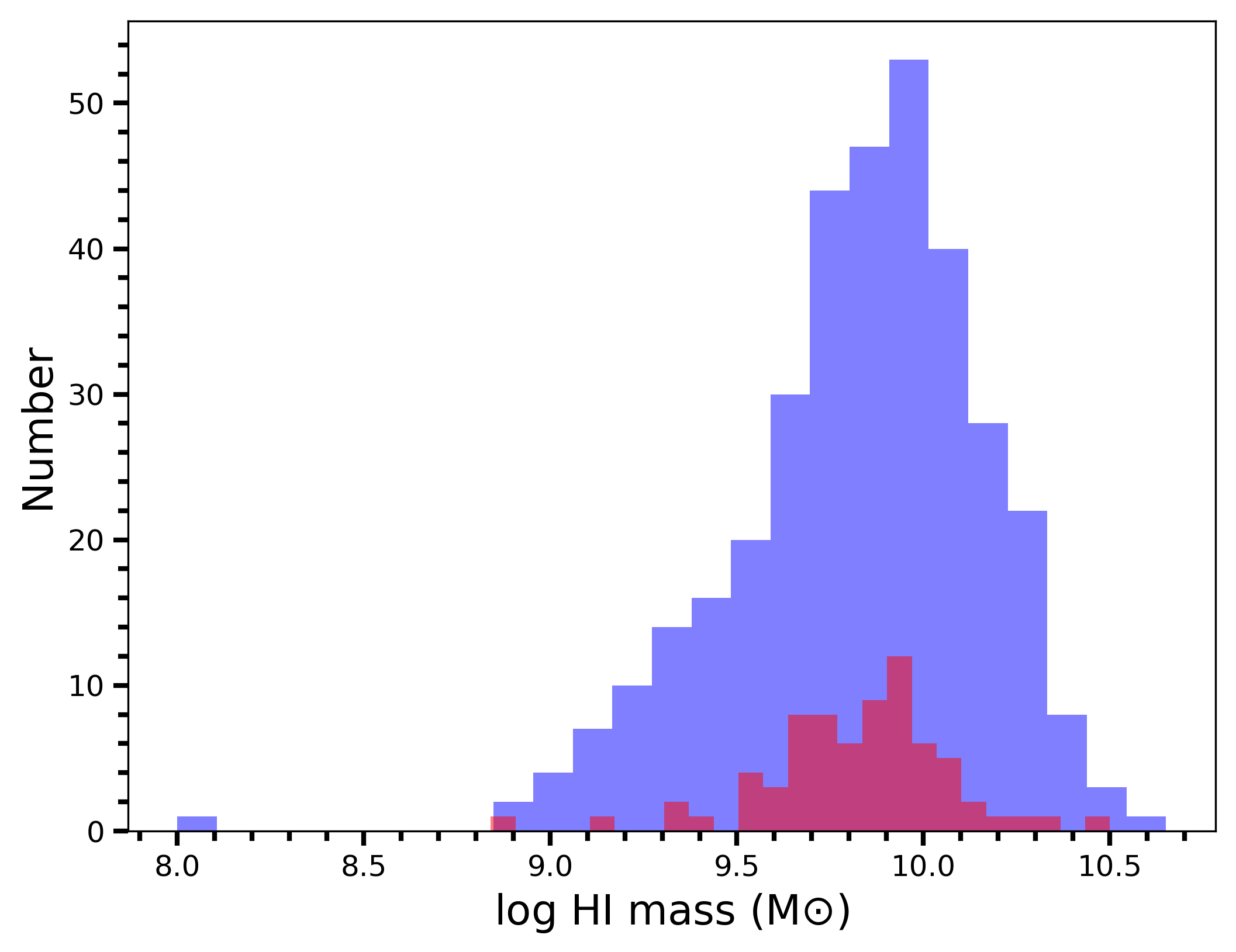}
\caption{Distribution of the integrated HI mass content of 422 barred galaxies. The combined distribution of starburst and main sequence galaxies is shown in blue. The passive galaxies are shown in red.}\label{figure:fig1}
\end{figure}

The quenched and star-forming population of barred galaxies, however, do not show a significant difference in HI content. The integrated HI mass (M$_{HI}$) is above 10$^9$ M$\odot$ for the quenched galaxies and we found that out of 422 galaxies with HI detection, 158 (37.44$\%$) are in starburst, 192 (45.5$\%$) in main sequence, and 72 (17.06$\%$) in the quenched regions of the SFR-M$\star$ plane. The distributions of the HI mass content of galaxies on the main sequence and starburst (shown in blue) along with the passive ones (shown in red) are shown in Fig. \ref{figure:fig1}. This figure suggests that the passive galaxies with HI detection have a similar gas content to that of the star-forming galaxies. The log M$_{HI}$/M$\star$ is $\sim$ -1.0 to -0.5 for quenched and star-forming massive galaxies. This implies that for a good fraction of galaxies star formation quenching is not due to a lack of fuel for star formation, but could be due to processes that can stabilize the gas against collapse and inhibit star formation. As discussed earlier, the action of stellar bars can facilitate this suppression of star formation.

The offset of galaxies from the main sequence relation ($\Delta$ MS) can be quantified using the main sequence relation of \citet{Renzini_2015}. If bar quenching is the primary or dominant mechanism for the transformation of barred galaxies to the passive phase, we expect to see a correlation between the offset of the galaxy from the main sequence and the scaled bar length. Figure \ref{figure:fig0}b shows that the offset of galaxies from the main sequence has no dependence with the average scaled length of the bar. The dotted line  signifies the region of main sequence, with galaxies outside of this region being either starburst or quenched. The galaxies are shown in color with the color scale set by the $D4000n$ strength. There is a dependence on the $D4000n$ strength as expected for quenched galaxies. The galaxies on and above the main sequence relation have $D4000n$ strengths lower than 1.5 (therefore stellar population ages < 1 Gyr). The galaxies below the main sequence relation with  $D4000n$ greater than 1.5 can be considered as hosting older stellar populations (no recent star formation) \citep{Kauffmann_2003b}. These galaxies are hence the quenched population of barred disk galaxies that are already on the passive population or in transition to join the passive galaxies. In order to further explore the dependence of the length of the bar on the star formation properties, we plotted the specific star formation rate (sSFR) against the stellar mass, color coding the points with respect to the scaled length of the bar. Figure \ref{figure:fig0}c demonstrates the sSFR-M$\star$ plane of the barred galaxies used in the present study as a function of the scaled bar length. We note that in the quenched population (log(sSFR) $<$ -11.5), bar length decreases with the stellar mass of the galaxy.

\section{Discussion}

\begin{table*}
\centering
\label{Quantifying bar quenching}
%\tablewidth{0pt}
\tabcolsep=0.6cm
%\tiny
\begin{tabular}{cllll} %23
\hline
Galaxy & Main sequence & Starburst & Quenched & Total\\
 \hline
%\\
M$\star$  > 10$^{11.0}$ M$\odot$   & 32~~~~~(11.55$\%$)   &  8~~~~~~ (2.88$\%$)   &   237~~~~(85.56$\%$) & 277\\
M$\star$  > 10$^{10.2}$ M$\odot$   & 551~~~(26.81$\%$)   & 138~~~(6.71$\%$)   &  1366~~(66.47$\%$) & 2055\\
M$\star$  < 10$^{10.2}$ M$\odot$  &  458~~~(55.18$\%$)   & 173~~~(20.84$\%$)  & 199~~~~(23.97$\%$) & 830\\
All\space                                          & 1009~(34.97$\%$)  & 311~~~(10.78$\%$)  & 1565~~(54.25$\%$) & 2885\\
%\\
\hline
%\hline
\end{tabular}
\caption{\label{t12} Number of barred galaxies separated into different mass bins and in different regions of the SFR-M$\star$ plane.}
\end{table*}

The importance of bar quenching for selected galaxies in the local Universe has been demonstrated in recent literature  \citep{James_2009,James_2015,James_2016,James_2018, George_2019,Newnham_2019}. The bar region of a few selected galaxies in the local Universe exhibits a central hole in the HI image \citep{Consolandi_2017,Newnham_2019}. Based on a multiwavelength analysis, the barred  galaxy Messier 95 is shown to be devoid of ongoing/recent star formation, as well as molecular and neutral hydrogen along the bar region. The galaxy can then follow an inside-out quenching scenario once the existing molecular hydrogen in the nuclear region is depleted \citep{George_2019}. The stellar bar in galaxies therefore can play a dominant role in regulating the gas supply and thereby suppress star formation at the central regions of disk galaxies, which, aided by other  secondary quenching mechanisms, can eventually lead to global quenching of star formation in these galaxies. The aim of this study is to understand the significance of bar quenching on the global quenching of barred galaxies.\\

As given in Table 1, 54.25$\%$ of local Universe barred galaxies used in the present study are quenched. The quenched fraction changes to 66.47$\%$ when galaxies with M$\star$ > 10$^{10.2}$ M$\odot$ are considered. This increases to 85.56$\%$ for the most massive galaxies in the sample (M$\star$ > 10$^{11}$ M$\odot$). These values suggest that a good fraction of  barred galaxies in the local Universe are quenched and the quenched fraction increases as a function of stellar mass. This can be due to bar quenching happening in massive barred disk galaxies. 
Longer bars can suppress star formation in a larger area in the central region of the galaxy and we expect to see maximum quenching for those galaxies with stronger stellar bars, if bar quenching has played a significant role in the global quenching of these galaxies. However, as shown in Fig. \ref{figure:fig0}b, there is no significant correlation between the length of the bar and the offset of the galaxy from the main sequence. This suggests that bar quenching may not have a significant role in the  transformation of barred galaxies into passive galaxies.\\

In Fig. \ref{figure:fig0}c we see that the sSFR decreases as a function of bar length on the main sequence and a reverse trend in the quenched population. A recent study  by \citet{Sheth_2008} showed that a significant fraction of massive barred galaxies were already assembled at high redshift (z $\sim$ 2). These galaxies could have quenched and evolved as passive galaxies in the local Universe. The quenched barred galaxies in the high mass end (M$\star$ > 10$^{10.5}$ M$\odot$)  may be the quenched descendants of the massive barred galaxies at higher redshifts. This could explain the weaker nature of the length of bars hosted by galaxies on the high mass end. The high mass galaxies might have been quenched by strong bars at z $\sim$ 2,  and over the timescale of $\sim$ 10 Gyrs these bars might have dissolved and become weaker or shorter. The low mass quenched galaxies might be recent arrivals and hence have long bars. This is in agreement with studies that suggest bars are dissolved or destroyed once the star formation is quenched  \citep{Raha_1991,Norman_1996,Martinez_2004,Shen_2004}. Some of the high redshift barred galaxies that are quenched and their bars dissolved/destroyed may not be identified as barred galaxies in morphology and hence may not be in the sample studied here. This could also explain the absence of correlation between the length of the bars and the offset from the star-forming main sequence of barred galaxies.\\

Recently, based on resolved HI imaging observations of six selected HI-rich barred galaxies, \citealt{Newnham_2019} demonstrated that the presence of a central HI hole depends on the dynamical age of the galaxy, and also that the possible presence of companions can alter the HI content of the barred galaxies. This implies that the environment can play a role in regulating the HI content, which can affect the star formation quenching of barred galaxies. This could be another reason for a non-correlation between the bar length and the position of the quenched galaxy on the SFR-M$\star$ plane. Environment can also affect the role and impact of secondary quenching mechanisms suggested for the suppression of star formation in the outer regions of the barred galaxies. A detailed environmental study of the barred galaxies can therefore bring more insights into discerning the significance of bar quenching on the global quenching of star formation in barred galaxies. However, \cite{guo_2019} suggest that massive galaxies ($>$ 10$^{10.5}$ M$\star$) are less likely to be affected by environmental effects and the majority of our sample of quenched barred galaxies are massive ($>$ 10$^{10.5}$ M$\star$).\\

Apart from dissolution of bars over time after quenching and  environmental effects, another reason for the absence of correlation between the length of the bar and the position of barred galaxies in the SFR-M$\star$ plane could be the action of other quenching mechanisms, which can be responsible for the quenching of star formation in the inner/central  regions of our sample galaxies. In our initial sample selection, we removed  galaxies that are classified as Seyfert galaxies based on emission line diagnostic diagrams. This suggests that the effect of quenching due to AGN feedback in our sample is negligible. However, the action of AGN feedback in the past cannot be ruled out. Many of the barred galaxies in our sample host bulges. We note that the bulge to total (B/T) ratio of barred galaxies in our sample that occupy  the passive region is higher than the ratio of those on the main sequence. This can be due to the enhancement of star formation at the central region of barred galaxies prior to bar quenching and buildup of pseudo bulges \citep{Kormendy_2004}.  Morphological quenching aided by bulges is generally driven by classical bulges, which are formed in major merger events. However, the coexistence of classical and pseudo bulges is possible and there could be some effects of morphological quenching in the presence of classical bulges.\\

Here we note one aspect of bar quenching demonstrated in simulations. All the recent simulations  \citep{Khoperskov_2018,Carles_2016} show a significant (a factor of ten) decrease in the star formation rate of barred galaxies on a timescale of $\sim$ 1 Gyr. In principle this is sufficient to shift the location of a galaxy from the main sequence to the passive region. However, all these simulations show that prior to the suppression of star formation all these galaxies were either in starburst phase or in the upper part of the main sequence in the SFR - M$\star$ plane. Thus the recent simulations are not able to demonstrate a case of bar quenching transforming a main sequence galaxy into passive galaxy. This is also true for the Milky Way galaxy, where a reduction in star formation rate by a factor of ten occurred almost 9 Gyr, which could be possibly accompanied by the formation of the stellar bar \citep{Haywood_2016}. It should be noted that the Galaxy is on the main sequence relation and certainly the action of the stellar bar only changes the position from a starburst to a more moderate case of star formation \citep{Renzini_2015}. We stress here that the scenario explored in this paper is that of the action of stellar bars being the dominant or primary quenching mechanism responsible for the transition of star-forming barred galaxies in the main sequence to the passive region.

\section{Summary}

 We quantified the fraction of quenched barred galaxies based on the position of galaxies on the SFR-M$\star$ plane. We demonstrate that in the local Universe 54.25$\%$ of barred galaxies are quenched, 34.97$\%$ are on the main sequence, and 10.78$\%$ are on the starburst phase. 
The quenched fraction becomes 66.47$\%$ for galaxies with M$\star$  > 10$^{10.2}$ M$\odot$, and  85.56$\%$ for the most massive galaxies in the sample (M$\star$ > 10$^{11}$ M$\odot$). This implies that a significant fraction of massive barred galaxies in the local Universe are quenched. We note that significant neutral hydrogen with M$_{HI}$ > 10$^{9}$ M$\odot$ and log M$_{HI}$/M$\star$ $\sim$ -1.0 to -0.5 is detected in quenched galaxies. The offset of the quenched barred galaxies from the main sequence relation is not dependent on the length of the stellar bar. This implies that the bar quenching may not be contributing significantly to the global quenching of star formation in barred galaxies. However, this observed result could also be due to other factors such as the dissolution of bars over time after star formation quenching, the effect of other quenching processes acting simultaneously, and/or the effects of environment. We found that the bar length decreases as a function of stellar mass for the quenched galaxies in the sSFR-M$\star$ plane. Future structural, stellar population, and environmental studies on quenched, massive barred galaxies as a function of redshift are necessary to better understand the present results from the local universe and the role of bar quenching in the global quenching of star formation in galaxies.

\begin{acknowledgements}

We thank the anonymous referee for the comments, which improved the scientific content of the paper. SS acknowledges support from the Science and Engineering Research Board, India, through the Ramanujan Fellowship.

\end{acknowledgements}

%-------------------------------------------------------------------

\end{document}